\documentclass[a4paper,11pt]{article}
\usepackage{jcappub} 

\usepackage{amsmath,amssymb,amsfonts}
\usepackage{color}
\usepackage{graphicx}
\usepackage{enumerate} 
\usepackage{colordvi} 
\usepackage{bm}
\usepackage{tensor}

\newcommand{\ve}{\mathbf{e}}

\newcommand{\vx}{\mathbf{x}}
\newcommand{\vy}{\mathbf{y}}
\newcommand{\vk}{\mathbf{k}}
\newcommand{\vz}{\mathbf{z}}

\newcommand{\wt}{\widetilde}

\newcommand{\etavk} {(\eta,\vk)}
\newcommand{\etavkp}{(\eta,\vk')}
\newcommand{\etapvkp}{(\eta',\vk')}
\newcommand{\etavx} {(\eta,\vx)}

\newcommand{\etapvxp}{(\eta',\vx')}
\newcommand{\etak} {(\eta,k)}
\newcommand{\etapk} {(\eta',k)}
\newcommand{\etaetapk} {(\eta,\eta',k)}
\newcommand{\etaetapvk} {(\eta,\eta',\vk)}

\newcommand{\be}{\begin{equation}}
\newcommand{\ee}{\end{equation}}
\newcommand{\bey}{\begin{eqnarray}}
\newcommand{\eey}{\end{eqnarray}}

\newcommand{\ttt} {t} 
\newcommand{\sss} {s} 
\newcommand{\HH} {H}  
\newcommand{\hh} {h}  
\newcommand{\ff} {f}  
\newcommand{\GG} {\gamma}  
\newcommand{\ag}  {{}^*\GG}
\newcommand{\at}  {{}^*\ttt}
\newcommand{\as}  {{}^*\sss}
\newcommand{\ah}  {{}^*\hh}
\newcommand{\af}  {{}^*\ff}

\newcommand{\aX}  {{}^*X}
\newcommand{\calH}{\mathcal{H}}

\newcommand{\nn}  {\nonumber}

\newcommand{\deltakk}  {(2\pi)^3\delta_D(\vk+\vk')}

\title{
Extracting parity-violating gravitational waves from projected tidal force tensor in three dimensions
}

\newcommand{\ASIAA}{Institute of Astronomy and Astrophysics, Academia Sinica,
No. 1, Section 4, Roosevelt Road, Taipei 10617, Taiwan}
\newcommand{\IPMU}{Kavli Institute for the Physics and Mathematics of the Universe (WPI), UTIAS, The University of Tokyo, Kashiwa, Chiba 277-8583, Japan}
\newcommand{\YITP}{Center for Gravitational Physics, Yukawa Institute for Theoretical Physics, Kyoto University, Kyoto 606-8502, Japan}
\newcommand{\LeCosPA}{Leung Center for Cosmology and Particle Astrophysics, National Taiwan University, Taipei 10617, Taiwan}

\author[a,b]{Teppei Okumura}
\author[b,c,d]{and Misao Sasaki}

\affiliation[a]{\ASIAA}
\affiliation[b]{\IPMU}
\affiliation[c]{\YITP}
\affiliation[d]{\LeCosPA}
\emailAdd{tokumura@asiaa.sinica.edu.tw}
\emailAdd{misao.sasaki@ipmu.jp}

\abstract{ Gravitational waves (GWs) may be produced by various
  mechanisms in the early universe.  In particular, if parity is
  violated, it may lead to the production of parity-violating GWs. In
  this paper, we focus on GWs on the scale of the large-scale
  structure. Since GWs induce tidal deformations of the shape of
  galaxies, one can extract such GW signals by observing images of
  galaxies in galaxy surveys.  Conventionally the detection of such
  signals is discussed by considering the three-dimensional power
  spectra of the $E/B$-modes.  Here, we develop a complementary new
  technique to estimate the contribution of GWs to the tidal force
  tensor field projected on the celestial sphere, which is a directly
  observable quantity.  We introduce two two-dimensional vector fields
  constructed by taking the divergence and curl of the projected tidal
  field in three dimensions.  Their auto-correlation functions
  naturally contain contributions of the scalar-type tidal field.
  However, we find that the divergence of the curl of the projected
  tidal field, which is a pseudo-scalar quantity, is free from the
  scalar contribution and thus enables us to extract GW signals.  We
  also find that we can detect parity-violating signals in the GWs by
  observing the nonzero cross-correlation between the divergence of
  the projected tidal field and the curl of it.  It roughly
  corresponds to measuring the cross-power spectrum of $E$ and
  $B$-modes, but these are complementary to each other in the sense
  that our estimator can be naturally defined locally in position
  space.  Finally we present expressions of the correlation functions
  in the form of Fourier integrals, and discuss the properties of the
  kernels specific to the GW case, which we call the overlap reduction
  function, borrowing the terminology used in the pulsar timing array
  experiments.  }
\arxivnumber{2405.04210}
\begin{document}

{\baselineskip0pt
\rightline{\baselineskip16pt\rm\vbox to20pt{
            \hbox{YITP-24-52}
\vss}}%
}

\maketitle

\flushbottom
\section{Introduction}\label{sec:intro}

Detecting cosmological gravitational wave (GW) backgrounds has become
one of the hot topics in cosmology. On very large scales, $\sim
\rm{Gpc}$, the primordial GWs originating from the vacuum fluctuations
during inflation \cite{Starobinsky:1979} can be probed by measuring
the $B$-mode signal in the polarization of the cosmic microwave
background (CMB) \cite{Zaldarriaga:1997,Kamionkowski:1997}.  On very
small scales, $\sim10^{8}-10^{13}\rm{cm}$ ($\sim
10^{-3}-10^2\rm{Hz}$), interferometric gravitational wave detectors,
either ground-based or space-based, may be able to detect a
cosmological GW background in the near future
\cite{Luo:2016,LISA:2017,LVK:2018,Kawamura:2021}.  Recently, several
pulsar time array (PTA) collaborations announced the detection of a GW
background on pc scales ($\sim 10^{-8}\rm{Hz}$)
\cite{NANOGrav:2023,EPTA_Collaboration_and_InPTA_Collaboration:2023,Xu:2023a}.
On the other hand, a vast range of wavelengths corresponding to the
large-scale structure of the universe, $\sim1-100\,\rm{Mpc}$, has not
been explored well, though a number of theoretical considerations have
been given, as GWs on these scales may be encoded in the large-scale
structure as dynamical and projection effects on the galaxy clustering
and weak gravitational lensing by GWs
\cite{Kaiser:1997,Dodelson:2003a,Cooray:2005,Yoo:2009,Masui:2010,Dodelson:2010,Jeong:2012a,Schmidt:2012,Schmidt:2012a,Chisari:2014}.

In all these studies, one of the intriguing questions asked is if
there exists parity violation in a GW background.  There is a growing
interest in finding parity asymmetry in cosmological observations,
e.g., birefringence
\cite{Minami:2020,Eskilt:2022,Diego-Palazuelos:2022}, galaxy spins
\cite{Iye:2019,Yu:2020,Motloch:2021,Motloch:2022,Shim:2024}, and
galaxy/CMB four-point correlation functions
\cite{Philcox:2022a,Hou:2023,Philcox:2023,Philcox:2024a}.  If parity
is violated in the early universe, it may source the tensor
perturbation of the metric, leading to the production of
parity-violating GWs
\cite{Lue:1999,Jackiw:2003,Seto:2006,Saito:2007,Seto:2007,Seto:2008,Satoh:2008,Jeong:2012,Maleknejad:2013,Masui:2017,Bastero-Gil:2023,Komatsu:2022}.
It was pointed out that astrometry can be used to detect parity
violation on pc scales \cite{Liang:2024}.  In this paper, we focus
on GWs with wavelengths $\sim 10-100\,\rm{Mpc}$.

The galaxy shapes are intrinsically aligned with the surrounding
large-scale dark matter distribution, known as intrinsic alignment
(IA) \cite{Lee:2000,Catelan:2001,Crittenden:2001,Hirata:2004}.
Orientations of the major axis of galaxies were shown to be linearly
aligned with the surrounding tidal field in both simulations and
observations \cite{Okumura:2009,Okumura:2009a,Blazek:2011}, which led
to the possibility of IA of galaxy shapes being a powerful probe of
various cosmological effects
\cite{Chisari:2013,Lee:2023,Schmidt:2015,Chisari:2016,Kogai:2021}.
Recent studies have developed accurate theoretical models for
statistics of galaxy IA in full three dimensions
\cite{Okumura:2019,Okumura:2020,Okumura:2020a,Akitsu:2021a,Shiraishi:2021a,Akitsu:2021,Shi:2021,Kurita:2022}.
Thus, the cosmological information encoded in the IA of galaxies can
be maximized by measuring galaxy shapes in galaxy redshift surveys
\cite{Taruya:2020,Okumura:2022,Chuang:2022,Saga:2023,Shiraishi:2023,
  Okumura:2023,Kurita:2023}.

Ref.~\cite{Schmidt:2014} proposed that GWs directly affect shapes of
galaxies as the instantaneous response effect, similarly to the scalar
tidal field case above.  Ref.~\cite{Biagetti:2020} extended the work
and investigated the possible upper limit on the parity-violating GWs,
but they still considered angular statistics.  Ref.~\cite{Akitsu:2023}
utilized a separate universe technique and confirmed using $N$-body
simulations the ansatz between the effects of GWs and scalar tidal
field on galaxy shapes proposed by Ref.~\cite{Schmidt:2014}.
Refs.~\cite{Akitsu:2023,Philcox:2024,Saga:2024} computed the
auto-power spectra of the $E$- and $B$-modes as well as their
cross-power spectrum in three dimensions.  The latter can be
particularly used to search for the parity-violation signals.  These
three-dimensional power spectra are observables in galaxy redshift
surveys, since the projected shapes of galaxies on the celestial
sphere are measured via imaging observations and the distances to them
are via spectroscopic observations
\cite{Okumura:2019,Taruya:2020,Kurita:2021,Matsubara:2024b}.  This is
opposed to the power spectra of CMB which provides angular
(two-dimensional) modes.  Furthermore, a galaxy survey targets
different scales, and thus it is complimentary to the CMB surveys.
One difficulty of measuring the effect of GWs on galaxy shapes in this
way is that it requires a large survey volume because the
$E$-/$B$-modes are quantities defined in Fourier space hence global in
real space. Thus, it is not straightforward to estimate the effects of
GWs locally in a small region of the universe.

In this paper, we develop and present a new technique to extract
gravitational wave signals from the projected tidal force field in
three dimensions.  We introduce two two-dimensional vector fields
constructed by taking the divergence and curl of the projected tidal
field. We show that taking their auto-correlations enables us to
extract signals of the GWs.  Furthermore, we show that
cross-correlating the divergence of the projected tidal field and the
curl of it allows us to extract parity-violating signals in the GWs.
These new estimators are complementary to the power spectra of the
$E$- and $B$-mode fields in the sense that our estimator can be
naturally defined locally in position space. 
To be more specific, the similarity is that both the EB decomposition used in the CMB community \cite{Kamionkowski:1997} and the method developed here focus on the tensor field projected on the celestial sphere. 
On the other hand, the main difference is that while the EB decomposition is defined in terms of spherical harmonics, hence globally defined from the real space point of view, our method involves only spatial derivative operations to single out the tensor components, hence is defined locally in space.
It is thus
straightforward to present the expressions of the correlation
functions.  Our formalism is admittedly quite fundamental in the sense
that it assumes an ideal observational situation. Nevertheless, we
hope it will eventually lead to a unique method for detecting an
extremely low frequency GW background, with the corresponding
wavelengths of 10 - 100 Mpc, that may or may not contain
parity-violating components.

The structure of the paper is as follows.  In section
\ref{sec:gw_preliminaries}, we give a review on the decomposition of
GWs into the polarizations and power spectra.  In section
\ref{sec:projection}, we present the formalism for extracting the GW
power spectrum from the projected tidal force field in three
dimensions.  Our conclusions are given in section
\ref{sec:conclusions}.  Appendix \ref{sec:EB} provides the relation
between our formalism and the E-/B-mode decomposition.  In Appendix
\ref{sec:power_w_trace} we present the power spectra of the projected
tidal field before the trace part is subtracted.
In Appendix \ref{sec:derivation}, we show detailed derivations for the formulas of the power spectra given in section \ref{sec:projection}. 

Throughout this paper, we use the Einstein summation convention, i.e.,
the summation is assumed when the same letters appear in the upper and
lower indices simultaneously.

\section{Gravitational waves in tidal force field}\label{sec:gw_preliminaries}

Let us start by defining GWs in the cosmological background. We define
GWs by the transverse and traceless tensor perturbation, $h_{ij}^{\rm
  TT}(\eta,\vx)$, of the metric. Together with the scalar
perturbation, the perturbed metric is expressed as
\begin{align}
ds^2 &=  -a^2(\eta)\Bigl([1+2\Psi(\eta,\vx)]d\eta^2 
+\big\{[1-2\Psi(\eta,\vx)]\delta_{ij} + h_{ij}^{\rm TT}(\eta,\vx) \big\}dx^idx^j\Bigr), \label{eq:metric_tt}
\end{align}
where $\eta$ is the conformal time, $\Psi(\eta,\vx)$ the Newton
potential, $a(\eta)$ the scale factor and $\delta_{ij}$ is Kronecker's
delta, and we assumed that the anisotropic stress is negligible.  In
the following, we omit the superscript ``TT'' from $h_{ij}^{\rm TT}$.

The tidal force is described by the geodesic deviation equation, with
the force given in terms of the Riemann tensor. At linear order in
cosmological perturbation theory, it is expressed in terms of the
perturbative components $\delta R^{k}{}_{0j0}$. For the metric
\eqref{eq:metric_tt}, it is given by
\begin{align}
\delta R^k{}_{0j0}\etavx=&
\left[\Psi''\etavx+2\calH(\eta)\Psi'\etavx+\frac{1}{3}\nabla^2\Psi\etavx\right]\delta^k_{j}
+\delta^{ki}F_{ij}\etavx\,,
 \label{eq:riemann}
\end{align}
where $\mathcal{H}\equiv a H$ is the conformal Hubble parameter and
$F_{ij}\etavx$ is the traceless part that defines the tidal field and
given by
\begin{equation}
F_{ij}\etavx\equiv
\left(\partial_i\partial_j-\frac{1}{3}\delta_{ij}\nabla^2\right)\Psi\etavx
-\frac{1}{2}\left[h_{ij}''\etavx+\calH(\eta) h_{ij}'\etavx\right]\,.
\label{eq:tidaldef}
\end{equation}
Below we focus on the traceless components.  To evaluate the
cosmological tidal effect, it is convenient to introduce a
dimensionless tidal field, defined by
\begin{align}
f_{ij}\etavx\equiv\frac{F_{ij}\etavx}{4\pi G\bar\rho (\eta)a^2}=\sss_{ij}\etavx+t_{ij}\etavx\,,
\label{eq:fijdef}
\end{align}
where
\begin{align}
\sss_{ij}(\eta,\vx) &\equiv \frac{1}{4\pi G\bar{\rho}a^2} \left( \partial_i\partial_j - \frac{1}{3}\delta_{ij}\nabla^2 \right) \Psi(\eta,\vx) 
=\left(\frac{\partial_i\partial_j}{\nabla^{2}} - \frac{1}{3}\delta_{ij}\right)
\delta_m(\eta,\vx),
\label{eq:ssijdef}\\
t_{ij}\etavx &\equiv -\frac{1}{8\pi G\bar{\rho}a^2}
\left[h_{ij}''\etavx+\calH(\eta) h_{ij}'\etavx\right]\,,
\label{eq:tijdef}
\end{align}
with $\bar\rho$ being the background energy density and
$\delta_m(\eta,\vx)$ the matter density contrast.  Note that we have
assumed the scale of our interest to be small enough so that the
Newtonian approximation is valid. In the real Universe, this should be
valid on scales $\lesssim500\,{\rm Mpc}$.

Due to the scalar potential term, it is hard to extract the tensor
part, GWs, from observations because it is subdominant, as estimated
below.  One of the goals of this paper is to formulate a method to
extract the tensor part from the tidal field based solely on
observables. As we intend to develop the basic formalism, in this
paper we assume an ideal situation for observation and ignore all
possible observational errors.

\subsection{Polarizations of gravitational waves}

We consider the tidal field in Fourier space, $X_{ij}(\eta,\vk) =
\int{d^3x}\,X_{ij}(\eta,\vx) e^{-i\vk\cdot\vx}$ ($X=\{s,t\,{\rm or}\, h \}$). The scalar contribution, $s_{ij}\etavk$, is given by $s_{ij}\etavk=\left(\hat k_i\hat k_j-(1/3)\delta_{ij}\right)\delta_m\etavk$.
We decompose GWs in
Fourier space into the right-handed ($R$) and left-handed ($L$)
polarizations (e.g., \cite{Misner:1973,Isi:2023}),
\begin{align}
h_{ij}(\eta,\vk) &= e_{ij}^{(R)}(\hat{\vk})h_{(R)}(\eta,\vk) + e_{ij}^{(L)}(\hat{\vk})h_{(L)}(\eta,\vk), \label{eq:hij} \\
e^{(R,L)}_{ij}(\hat{\vk}) & =\frac{1}{\sqrt{2}}\left(e^{(+)}_{ij}(\hat{\vk})\pm i \, e^{(\times)}_{ij} (\hat{\vk})\right), 
\end{align}
where hat denotes a unit vector, for instance $\hat{\vk}=\vk/k$, and
$e^{(+)}_{ij}  = \frac{1}{\sqrt{2}} \bigl( e^{(1)}_i e^{(1)}_j  - e^{(2)}_i e^{(2)}_j \bigr)$ and 
$e^{(\times)}_{ij} = \frac{1}{\sqrt{2}} \bigl( e^{(1)}_i e^{(2)}_j  + e^{(2)}_i e^{(1)}_j \bigr)$,
with $\ve^{(1)}$ and $\ve^{(2)}$ being arbitrary orthonormal vectors
spanning two-dimensional space orthogonal to $\hat{\vk}$. We assume
the set ($\ve^{(1)}$, $\ve^{(2)}$, $\hat{\vk}$) forms a right-handed
Cartesian basis.  All the polarization tensors are defined to be
orthonormal, $e_{ij}^{(p)} e_{(p')}^{ij} = \delta^{p}_{p'}$ for
$p,p'=\{+, \times\}$, and $e_{ij}^{(p)}\,\bar{e}_{(p')}^{ij} =
\delta^{p}_{p'}$ for $p,p'=\{R, L\}$, where a bar stands for the
complex conjugate.

In passing, we note that it is useful to express the Fourier component
of the dimensionless tidal field $t_{ij}$ given by \eqref{eq:tijdef}
in terms of $h_{ij}$.  Assuming $k\gg\calH$, we have $t_{ij}\propto
h_{ij}''=-k^2h_{ij}$. Hence
\begin{eqnarray}
t_{ij}\etavk=\frac{k^2}{3\calH^2}h_{ij}\etavk\,,
\label{eq:t=h}
\end{eqnarray}
where we have used the background Friedmann equation for a spatially
flat universe, $8\pi G\bar{\rho}a^2=3\calH^2$.  Noting that
$s_{ij}=O(\delta_m)$, the tensor contribution can be comparable to the
scalar one only when $h_{ij}\sim(\calH^2/k^2)\delta_m$. On 10 Mpc
scale, which is approximately the scale of our interest, we have
$\delta_m=O(1)$.  This means we need $h_{ij}\sim 10^{-4}$ to dominate
over the scalar contribution on that scale, which seems practically
impossible to attain.  Conversely, we need a method to cleanly
separate out the scalar contribution if we are to detect the tensor
part in the tidal field.

We now define the symmetric curl field of a given tensor field
$X_{ij}$ \cite{Creminelli:2014} by putting the asterisk on the
left-hand side of the symbol, ${}^* X_{ij}$, as
\begin{align}
{}^* X_{ij}(\eta,\vx) &\equiv 
\frac{1}{2}
\left( 
\epsilon\indices{_i^{mn}}\partial_m X_{nj} + \epsilon\indices{_j^{mn}}\partial_m X_{ni} 
\right)\,, \label{eq:curl_Xij}
\end{align}
where $\epsilon_i{}^{mn}=\delta_{ij}\epsilon^{jmn}$ and
$\epsilon^{jmn}$ is the totally anti-symmetric unit tensor with
$\epsilon^{123}=1$.  By taking the spatial derivative of
Eq.~(\ref{eq:hij}), we get
\begin{align}
\partial_m h_{nj}\etavx =
\int \frac{d^3k}{(2\pi)^3} & i \, k_m\left[ 
 e_{nj}^{(R)}(\hat{\vk})h_{(R)}\etavk 
+ 
e_{nj}^{(L)}(\hat{\vk})h_{(L)}\etavk \right] e^{i\vk\cdot\vx}. \label{eq:deriv_hij}
\end{align}
Then the curl of the tidal field reads
\begin{align}
\ah_{ij}(\eta,\vx) 
= \int\frac{d^3k}{(2\pi)^3} \ah_{ij}(\eta,\vk) e^{i\vk\cdot\vx}, 
\end{align}
where 
\be
\ah_{ij}(\eta,\vk) = k\left[ e_{ij}^{(R)}(\hat{\vk}) h_{(R)} (\eta,\vk)- e_{ij}^{(L)}(\hat{\vk}) h_{(L)}(\eta,\vk) \right]. \label{eq:curl_hij}
\ee
In the above we used the identities,
$i\,\epsilon\indices{_i^{mn}} k_m e_{nj}^{(R,L)} = \pm \, k\,e^{(R,L)}_{ij}$
\cite{Alexander:2005}.
Note that apart from the flip of sign of the L-mode, $\ah_{ij}$ is
essentially equivalent to the spatial derivative of $h_{ij}$.  Since
the curl of the scalar component is identically zero, the above
equation states that if we detect the non-zero curl component of the
tidal force field at large scales where linearized theory give an
accurate prediction, it provides a direct evidence of the signal of
gravitational waves.

\subsection{Power spectra of tidal force field}\label{sec:power}

The auto-power spectra of the gravitational waves $\hh_{ij}$ and their curl field $\ah_{ij}$ are given by
\begin{align}
\big\langle h^{ij}\etavk h_{ij}\etapvkp \big\rangle 
&=
k^{-2}\big\langle \ah^{ij}\etavk \ah_{ij}\etapvkp \big\rangle 
\nn \\ &
=
\deltakk
P_h \etaetapk~,\label{eq:hijhij} 
\end{align}
where $P_h(\eta,\eta',k)$ is the temporal auto correlation power spectrum that reduces to the standard power spectrum at equal time $\eta=\eta'$.
The (dimensionless) tidal field $f_{ij}$ consists of the scalar and
tensor components as given by \eqref{eq:fijdef}.  We mention again
that since the scalar component is curl-free, $\as^{ij}=0$, measuring
$^*f_{ij}$ enables us to extract the signal of GWs $t_{ij}$ out of the
tidal field $f_{ij}$.

The power spectra of each polarization is defined similarly as
\begin{align}
\big\langle \hh^{(\lambda)}\etavk\hh_{(\lambda)}\etapvkp \big\rangle 
&=k^{-2}\big\langle \ah^{(\lambda)}\etavk\ah_{(\lambda)}\etapvkp \big\rangle
\nn \\ &
=\deltakk
P_{(\lambda)}\etaetapk,\label{eq:hRLhRL}
\end{align}
where $\lambda = \{R,L\}$ and $P_{(R)} = P_{(L)} = P_h/2$ for
unpolarized GWs.  The cross-power spectra of the curl field $\ah_{ij}$
and the original field $\hh_{ij}$ are given by
\begin{align}
\big\langle \ah^{ij}\etavk\hh_{ij}\etapvkp \big\rangle 
&=\deltakk k\, \chi\etaetapk P_h\etaetapk, \label{eq:tijatij} \\
\big\langle \ah^{(\lambda)}\etavk\hh_{(\lambda)}\etapvkp \big\rangle 
&=\deltakk
k\,\chi\etaetapk P_{(\lambda)}\etaetapk,
\end{align}
where we introduced the chiral parameter $\chi(\eta,k)$ to quantify
parity violation in GWs,
\begin{align}
\label{chiral}
\chi\etaetapk\equiv \frac{P_{(R)}\etaetapk-P_{(L)}\etaetapk}{P_h\etaetapk } ~.
\end{align}
Namely, taking the cross correlation enables us to extract a signal of
parity violation \cite{Masui:2017}.

Let us now turn to the general properties of the power spectrum. First
we formally write the total tensor field,
$\hh_{(\lambda)}^{(\mathrm{total})}$, as a sum of the contributions of
primordial GWs, denoted by $\hh_{(\lambda)}$, and that generated by a
source $S_{(\lambda)}$ at some epoch in the early universe, denoted by
$\HH_{(\lambda)}$. Thus
$\hh_{(\lambda)}^{(\mathrm{total})}=\hh_{(\lambda)}+\HH_{(\lambda)}$,
where $\Box\hh_{(\lambda)}=0$ and $\Box\HH_{(\lambda)}=S_{(\lambda)}$.
To compute $\HH_{(\lambda)}$, we assume the wavenumbers are much
larger than the Hubble parameter at the time of generation,
$k\gg\calH$, which is reasonable from causality.  Let us introduce
$g_{(\lambda)}\etavk$ by
$\HH_{(\lambda)}\etavk=g_{(\lambda)}\etavk/a(\eta)$. The green
function at the $k\eta\gg 1$ limit is given by (e.g.,
\cite{Maggiore:2018})
\begin{align}
G(\eta-\eta')=\Theta(\eta-\eta')\sin{[k(\eta-\eta')]}/k,
\end{align}
where $\Theta(\eta)$ is the step function.  Assuming that GWs are
generated during an epoch, $\eta_i < \eta < \eta_i+\Delta\eta$, the
solution $g_{(\lambda)}$ is given by
\begin{align}
g_{(\lambda)}\etavk = 
\int_{\eta_i}^{\eta_i+\Delta\eta} d\eta' \frac{\sin{[k(\eta-\eta')]}}{k}a(\eta')S_{(\lambda)}(\eta',\vk),
\end{align}
where $S_{(\lambda)}(\eta,\vk)$ is the Fourier component of
$S_{(\lambda)}(\eta,\vx)$.  Regardless of the form of $S_{(\lambda)}$,
$g_{(\lambda)}$ has the form of $g_{(\lambda)}\etavk = g_0\etavk
\cos{(k\eta+\phi_\vk)}$ at $\eta > \eta_i+\Delta\eta$.  Taking the
product with itself and ensemble average, the $\phi_\vk$ term vanishes
and we obtain
\begin{align}
P_{(\lambda)}^{S}\etaetapk
= \frac{\big\langle |g_0(k)|^2 \big\rangle}{a(\eta)a(\eta')}\cos{[k(\eta-\eta')]}, \label{eq:pk_green}
\end{align}
where the superscript $S$ indicates its the spectrum of GWs generated
by a source.  As for the primordial GWs, the corresponding power
spectrum can be expressed as
\begin{align}
P_{(\lambda)}^{P}\etaetapk = T\etak T\etapk P_{(\lambda)}^{\rm ini}(k),
\label{eq:PrimP}
\end{align}
where the superscript $P$ denotes its primordial and $T\etak$ is the
transfer function of GWs. For $k\gg\calH$, we have
$T\etak\simeq\cos{[k\eta+\varphi_k]}/a(\eta)$ where $\varphi_k$ is not
random now but determined by the history of the equation of state of
the early universe (for example, if the universe were
radiation-dominated for the entire history,
$\varphi_k=0$). Nevertheless, taking into account the effect of
inhomogeneities in the universe, it is expected that the phase
$\phi_k$ will acquire an uncontrollable random noise after the wave
has propagated over a distance much larger than the wavelength
\cite{Allen:2000,Margalit:2020}. Therefore, for the modes of our interest,i.e., $k\eta\gg1$, the power spectrum $P_{(\lambda)}^P\etaetapk$
reduces to the form same as $P_{(\lambda)}^S\etaetapk$ in
\eqref{eq:pk_green}.

Thus the two-point correlation function can be generically expressed as
\begin{align}
\xi_{(\lambda)}(\eta,\eta',\vx-\vx')=
\int\frac{k^2dk}{(2\pi)^3}\int d\Omega_kP_{(\lambda)}\etaetapk
e^{i\vk\cdot(\vx-\vx')}\,,
\end{align}
where 
\begin{eqnarray}
P_{(\lambda)}\etaetapk
&=&P_{(\lambda)}^P\etaetapk+P_{(\lambda)}^S\etaetapk
\nonumber\\
&=&\frac{a_0^2}{a(\eta)a(\eta')}\cos[k(\eta-\eta')]\,P_{(\lambda)}(k)\,.
\end{eqnarray}
The $\cos[k(\eta-\eta')]/\bigl(a(\eta)a(\eta')\bigr)$ dependence of
the two-point function is the characteristic feature of cosmological
GWs, either primordial or generated by a transitional source, in
contrast to the case of the two-point function of the density
contrast, $\langle\delta_m(k,\eta)\delta_m(k,\eta')\rangle\propto
a(\eta)a(\eta')$.  Finally, we note that because the time dependence
of $P_{(\lambda)}$ is independent of the polarization $\lambda$, the
chiral parameter defined in \eqref{chiral} reduces to a function of
only $k$, $\chi=\chi(k)$.

\section{Projected tidal force field in three dimensions}\label{sec:projection}

In the previous section, we saw how the two-point correlation
functions of GWs behave and how the parity-violating component shows
up. However, they cannot be constructed solely from observables
because we can observe only the past lightcone, which is a
three-dimensional surface, while the quantities discussed in the
previous section assume that we can observe $h_{ij}$ in the whole
four-dimensional spacetime.  In this section, to overcome this
difficulty, we develop a formalism that uses only the observable
quantities.  Galaxy morphologies, such as the spins and ellipticities
of galaxies, are projected onto the celestial place in
observation. Analogously, we deal only with the tensor perturbation
projected on the celestial sphere, with the line of sight direction
identified as the time toward the past.

\begin{figure}[bt]
\begin{center}
\includegraphics[width=0.49\textwidth,angle=0,clip]{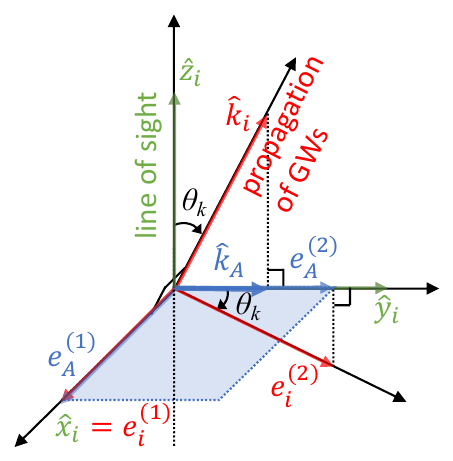}
\caption{ Illustration of the coordinates adopted in the text.  The
  direction of $\hat\vz$ is the observer's line of sight, and the
  orthonormal vectors $\hat\vx,\hat\vy$ and $\hat\vz$ form a
  right-handed Cartesian basis. The direction of propagation of a GW
  is specified by $\ve^{(3)} = \hat{\vk}$, tilted from the line of
  sight by an angle $\theta_k$, and $\ve^{(1)}, \ve^{(2)}$ and
  $\ve^{(3)}$ form another right-handed Cartesian basis. Here
  $\ve^{(1)}$ and $\hat\vx$ are chosen to coincide with each other,
  $\ve^{(1)}=\hat\vx$.  The components of the basis vectors projected
  onto the celestial sphere are denoted by capital letters such as
  $e^{(1)}_A$.  }
\label{fig:coordinate}
\end{center}
\end{figure}

For this purpose, in this section we need to be more specific about
the coordinates we adopt.  We choose the direction along observer's
line of sight as $\hat{\vz}$, where the hat means it is a unit
vector. Since the direction of propagation of GWs is arbitrary, we set
the angle between the direction and the line of sight to $\theta_k$
and the directional cosine, $\hat{\vk}\cdot
\hat{\vz}=\cos{\theta_k}\equiv\mu_k$.  The polarization tensor lies on
the surface perpendicular to $\hat{\vk}$, spanned by $\ve^{(1)}$ and
$\ve^{(2)}$.  The celestial sphere is locally spanned by two
orthonormal basis vectors, say, $\hat\vx'$ and $\hat\vy'$. In this
coordinate system, the components of $\hat{\vk}$ may be expressed as
$\hat{\vk}=(\sin\theta_k\cos\phi_k,\sin\theta_k\sin\phi_k,
\cos\theta_k)$. For convenience, one can rotate the axes by
$\pi/2-\phi_k$ to make the new basis vector $\hat{\vx}$ coincide with
$\ve^{(1)}$. Namely, we choose the $x$-axis such that
$\hat{\vx}=\ve^{(1)}$.  Then the components, $\ve^{(1)}$, $\ve^{(2)}$
and $\hat{\vk}$, respectively, have in the Cartesian coordinates,
\begin{align}
e_i^{(1)}=(1,0,0), \quad
e_i^{(2)}=(0,\cos{\theta_k},-\sin{\theta_k}), \quad
\hat{k}_i=(0,\sin{\theta_k},\cos{\theta_k}).
\end{align}
Then the components of the basis vectors defined/projected on the
celestial sphere in two dimensions are given by
\begin{align}
e_A^{(1)}=\hat{x}_A=(1,0), \quad
e_A^{(2)}=\hat{y}_A\cos{\theta_k}=(0,\cos{\theta_k}), \quad
\hat{k}_A=\hat{y}_A\sin{\theta_k}=(0,\sin{\theta_k}).
\end{align}
where the capital Latin indices denote quantities on the
two-dimensional celestial sphere.  Figure \ref{fig:coordinate} is an
illustration of the coordinate system.

Note that the tidal force field is characterized by the second time
derivative, $a^{-1}(a{h}_{ij}')'$, rather than $\hh_{ij}$ itself, as
shown in Eq.~(\ref{eq:riemann}). For the scale of our interest,
$k^2\gg\calH^2$, we can approximate it by $h_{ij}''$.  Thus, the power
spectra relevant to observations are those of the second time
derivative of GWs, and for instance the power spectrum of given fields
$X$ and $Y$, $\langle X^{ij} Y_{ij} \rangle=\deltakk P_{XY}(k)$,
should be replaced by $\langle \tilde{X}^{ij}
\tilde{Y}_{ij}\rangle=\deltakk \wt{P}_{XY}(k)$, where $\tilde{X}=X''$
and $\wt{P}_{XY}=k^4 P_{XY}(k)$.  In the following, however, we
consider $h_{ij}$ itself. This means all the resulting formulas should
be multiplied by $k^4$ when they are to be considered as the tidal
force field.

\subsection{Power spectra of projected tidal force field}\label{sec:power_projected}

Since only the tidal field projected on the sphere perpendicular to
the line of sight is observable, we consider the projected field of
GWs, $h_{AB}\etavx$ in real space or $h_{AB}\etavk$ in Fourier space.
Note that the vectors $\vx$ and $\vk$ are in three dimensions. We
assume that we can measure the three-dimensional positions of galaxies
via, say, measuring redshifts in spectroscopic galaxy surveys.  The
main purpose of this section is to derive formulas that can be used to
identify the GW contribution in the tidal field solely from the
projected field $h_{AB}$.

Let us first note that the tidal force field relevant to observables
is a traceless part in two dimensions,
\begin{align}
f_{AB}^T\etavx=\sss_{AB}^T\etavx+t_{AB}^T\etavx\,,
\label{eq:fTABdef}
\end{align}
where a quantity $X$ with the superscript $T$ stands for
$X_{AB}^T=X_{AB}-\frac12\delta_{AB}X\indices{^C_C}$.  In appendix
\ref{sec:power_w_trace} we present the power spectra of the projected
field before the trace part is subtracted.

It is straightforward to compute the power spectrum of the projected
scalar tidal field. 
The resulting spectrum is obtained as
\begin{align}
\left\langle \sss^{T,AB}\etavk\sss^T_{AB}\etapvkp\right\rangle
& =\frac12 \deltakk k^4 (1-\mu_k^2)^2 P_\psi \etaetapk, \label{eq:sTabsTab}
\end{align}
where $\mu_k=\hat{\vk}\cdot\hat{\vz}$ and
$\left\langle  \psi\etavk\psi\etapvkp\right\rangle=\deltakk P_\psi\etaetapk$ with $\psi\etavk\equiv\Psi\etavk/4\pi G\bar{\rho}a^2$. 
Let us then compute the power spectra of the projected GWs. 
The GW power spectrum is obtained as
\begin{align}
\left\langle \hh^{T,AB}\etavk\hh^T_{AB}\etapvkp\right\rangle
&=\frac18\deltakk(1+6\mu_k^2+\mu_k^4) P_h\etaetapk. \label{eq:hTabhTab}
\end{align}
Taking the average over $\mu_k$ leads to the monopole component,
$\frac{2}{5} P_h\etaetapk $.
However, it is hard to extract the GW signals via these expressions as
the scalar potential term, $\left\langle
\sss^{T,AB}\etavk\sss^T_{AB}\etapvkp\right\rangle$, dominates the
signal, which prevent us to extract out the GW contribution directly.
In practice the scalar contribution is the one commonly measured in
the analysis of intrinsic alignment of galaxy shapes (see discussion
in section \ref{sec:conclusions} below).  In appendix \ref{sec:EB} we
show how they are related to the power spectra of E- and B-modes.
The derivations of these equations are given in Appendix \ref{sec:derivation}.

\subsection{Power spectra of divergence and curl of projected tidal force field }\label{sec:power_div_curl}

We now consider to isolate the GW contribution from the observed projected tidal force field. 
We introduce the two-dimensional vector fields, $X_A$ and $^*X_A$, constructed respectively by taking the divergence and curl of the original projected field $X_{AB}\etavx$ ($X=\{s,t \ ({\rm or}\ h )\}$) as
\be
X_A\etavx \equiv \partial^B X_{BA}\etavx, \qquad 
\aX_A\etavx \equiv \epsilon^{BC}\partial_B X_{CA}\etavx , \label{eq:XA_aXA}
\ee
where $\epsilon^{BC}$ is the Levi-Civita symbol defined in two
dimensions.  Similarly, the two-dimensional vector fields after
subtracting the trace part, $X^T_A$ and $^*X^T_A$, are obtained from
$X^T_{AB}\etavx$ as
\be
X^T_A\etavx \equiv \partial^B X^T_{BA}\etavx, \qquad 
\aX^T_A\etavx \equiv \epsilon^{BC}\partial_B X^T_{CA}\etavx .\label{eq:XTA_aXTA}
\ee
Note that $\as_{A}=0$ but $\as^T_{A}\neq 0$. That is, unlike the
three-dimensional case, the curl of the scalar component does not
vanish unless one can accurately identify the trace part of $s_{AB}$.
Taking the curl of the projected tidal force field therefore does not
immediately remove the scalar contribution.

In Fourier space, the divergence and curl of the field $\ff_{AB}$ are
expressed as
\begin{align}
\ff^T_A\etavk &= i\, k^B \ff^T_{BA}\etavk = \ff_A\etavk+\frac{i}{2}k_A\left[ t\indices{^z_z}\etavk-k^Ck_C\,\psi\etavk \right],\\
\af^T_A\etavk &= i\, k_B \epsilon^{BC}\ff^T_{CA}\etavk=\af_A\etavk+\frac{i}{2}k_B\epsilon\indices{^B_A} \left[ t\indices{^z_z}\etavk+k^Ck_C\,\psi\etavk \right].
\end{align}
Nonzero power spectra for the divergence and curl of the projected
scalar tidal field with or without the trace part subtracted have the
same form, and are given by
\begin{align}
\left\langle \sss^{T,A}\etavk\sss^T_A\etapvkp \right\rangle  
& 
= \left\langle \as^{T,A}\etavk\as^T_A\etapvkp \right\rangle  
\nn \\& 
= 
\frac14 \deltakk
k^6 (1-\mu_k^2)^3 P_\psi\etaetapk \, . \label{eq:sTsT}
\end{align}
They are also related to the power spectrum of the original projected
scalar tidal field via a simple relation,
$\left\langle \sss^{T,A}\etavk\sss^T_A\etapvkp \right\rangle =\frac12 k^2 (1-\mu_k^2)\left\langle \sss^{T,AB}\etavk\sss^T_{AB}\etapvkp \right\rangle$.
Below we intend to formulate two-point statistics of $\at^T_A$, or
equivalently $\ah^T_A$ as given by \eqref{eq:t=h}.

We can also derive the auto-power spectra for the divergence and curl
of the traceless projected field.  They are given by
\begin{align}
\left\langle \hh^{T,A}\etavk\hh^T_A\etapvkp \right\rangle  
& 
= 
\left\langle \ah^{T,A}\etavk\ah^T_A\etapvkp \right\rangle  
\nn \\ & 
= 
\frac{1}{16}\deltakk
k^2(1-\mu_k^2)(1+6\mu_k^2+\mu_k^4) P_h\etaetapk \nn \\
&=\frac{1}{2}k^2(1-\mu_k^2)\left\langle \hh^{T,AB}\etavk\hh^T_{AB}\etapvkp \right\rangle  
. \label{eq:power_proj_auto_h_traceless}
\end{align}
By taking an average over angle $\mu_k$, we may extract out the
monopole component $\propto k^2 P_h(k)$.  We can similarly obtain the
quadrupole and hexadecapole moments.  As mentioned above and similarly
to equation (\ref{eq:hTabhTab}), this contribution of GWs cannot be
separated from that of the scalar potential term, $\left\langle
s^{T,A}\etavk s^T_A \etapvkp\right\rangle$.

Since $\as^{T,A}\propto\epsilon^{BA}\partial_B\psi$, we cannot isolate the GW signals by computing the auto correlation of the curl of the projected field. 
However, if we take the divergence of the projected field, the scalar component vanishes, $\partial_A\as^{T,A}\equiv 0$. Thus, similarly to the full three-dimensional case, Eq.~\eqref{eq:hijhij}, we can extract out GWs from the projected tidal force field. Specifically, let us introduce the symbols,
\begin{align}
\aX^T\etavx = \partial^A \, \aX^T_A\etavx, \qquad \aX^T\etavk = ik^A\,\aX^T_A\etavk.
\end{align}
By setting $X^T=h^T$, we obtain a direct probe of a GW signal. 
The power spectrum is given by
\begin{align}
\left\langle \ah^{T}\etavk\ah^T\etapvkp \right\rangle  
&
= 
\frac{1}{4}\deltakk
k^4\mu_k^2(1-\mu_k^2)^2 P_h\etaetapk . \label{eq:power_proj_auto_h_T}
\end{align}
We emphasize that measuring the above requires only the observable
quantities.

Now we consider the possibility to detect a signal of parity violation
from the projected two-dimensional field of GWs, similar to the
three-dimensional case given by Eq.~(\ref{eq:tijatij}).  Our basic
assumption is that we can only measure the projected tidal field, but
in three dimensions.  In section \ref{sec:power} we have seen that the
parity violation appears in the cross correlation of $h_{ij}$ and
$\ah_{ij}$. Our goal here is to derive a similar equation in terms of
the projected field. A straightforward choice for this purpose is to
consider the cross-correlation of $\hh^T_{A}$ and $\ah^T_{A}$.  We
obtain
\begin{align}
\left\langle \ah^{T,A}\etavk\hh^T_A\etapvkp \right\rangle  
& 
= 
\frac{i}{4}\deltakk
k^2\mu_k (1-\mu_k^4) \chi(k)P_h\etaetapk. \label{eq:power_proj_cros}
\end{align}
This is one of the main results of this paper.  This power spectrum
becomes nonzero if and only if parity is violated and the effect
appears in the odd multipole moments.  This cross power corresponds to
the cross power spectrum of the $E$- and $B$-modes. The relation to
the $E/B$ decomposition is discussed in Appendix \ref{sec:EB}.
The derivations of the expressions for all the power spectra for the scalar and tensor tidal fields in this subsection are provided in Appendices \ref{sec:derivation_scalar} and \ref{sec:derivation_tensor}, respectively.

\subsection{Correlation functions and overlap reduction functions}\label{sec:tpcf}

Here we perform the Fourier transform of the power spectra of the
projected tidal force field with the trace part subtracted, derived in
the previous subsection, and present the expressions of the
corresponding correlation functions. The two-point function of the
projected field $h^{T}_{AB}$ is given by
\be
\left\langle h^{T,AB}\etavx 
h_{AB}^{T}\etapvxp 
\right\rangle = \int \frac{k^2 dk}{(2\pi)^3}\int d\Omega_{k}P^T_{h\, h}\etaetapvk 
e^{i \vk\cdot(\vx-\vx')}, \label{eq:tpcf_XY}
\ee
where $P_{h\, h}^T$ is the power spectrum derived in the previous
subsection,
\begin{align}
P^T_{\hh\,\hh}\etaetapvk&=\frac{1}{8}
(1+6\mu_k^2+\mu_k^4) P_h\etaetapk \, .
\end{align}
To evaluate the two-point function, let us first specify the four
vectors of the two points, $x'^\alpha$ and $x^\alpha$, as
\begin{align}
& x'^\alpha = (\eta',x'^i)= (\eta_0-r',0,0,r'), 
\\ &
x^\alpha = (\eta,x^i)= (\eta_0-r, r\sin{\theta}\cos{\phi},r\sin{\theta}\sin{\phi},r\cos{\theta}),
\end{align}
where the observer's position is set to the origin, and $r=|\vx|$ and
$x^i=r\hat{x}^i$.  Also, we set the components of the four wavevector
$k_\alpha$ as
\be
k_\alpha=(-k, k\sin{\theta_k}\cos{\phi_k}, k\sin{\theta_k}\sin{\phi_k}, k\cos{\theta_k}),
\ee
where $k=|\vk|$. Here we consider a more general coordinate system
than that in Fig. \ref{fig:coordinate} in which the wavevector can
point to arbitrary directions.  We can then perform the $\phi_k$
integral in equation (\ref{eq:tpcf_XY}) to obtain
\be
\int^{2\pi}_0\frac{d\phi_k}{2\pi}e^{i\vk\cdot(\vx-\vx')}
=J_0(kr\sin{\theta}\sin{\theta_k})e^{ik(r'-r\cos{\theta})\cos{\theta_k}},
\ee
where $J_0$ is the $0$-th order Bessel function. 

To proceed further, we introduce $\Delta\eta\equiv \eta-\eta'=r'-r$.
Eventually, the correlation function is expressed as
\begin{align}
\left\langle \hh^{T,AB} \etavx \hh^T_{AB} \etapvxp
\right\rangle &= \frac{a_0^2}{a(\eta)a(\eta')}\int \frac{k^4 dk}{2\pi^2}P_h(k)\Gamma_{\hh\,\hh}(kr,k\Delta\eta,\theta).
\end{align}
Here, let us call the kernel $\Gamma_{h\, h}$ an overlap reduction
function (ORF), borrowing the terminology used in the context of
pulsar timing arrays \cite{Flanagan:1993,Allen:1999,Romano:2017} where the kernel for GWs is called the
Hellings-Downs curve \cite{Hellings:1983}. In our case, it is given by
\begin{align}
\Gamma_{\hh\,\hh}(kr,k\Delta\eta,\theta)&= 
\cos{(k\Delta\eta)}\int^1_0 d\mu_k\,\frac{1+6\mu_k^2+\mu_k^4}{8}J_0\left( A\sqrt{1-\mu_k^2}\right)\cos{\left(B\mu_k\right)}, \label{eq:HD_hh_hh}
\end{align}
where 
\begin{eqnarray}
 A(kr,\theta)=kr\sin{\theta}
 , \qquad 
B(kr,k\Delta\eta,\theta)&=&-k\Delta\eta+2kr\sin^2{(\theta/2)}\,.
\end{eqnarray}
We can derive the similar expression for the correlation function of
the projected scalar tidal field,
\begin{align}
&\left\langle \sss^{T,AB} \etavx
\sss^T_{AB} \etapvxp
\right\rangle = \int \frac{k^4 dk}{2\pi^2}P_\psi\etaetapk\Gamma_{\sss\,\sss}(kr,k\Delta\eta,\theta), 
\end{align}
where the scalar ORF is given by
\begin{align}
\Gamma_{\sss\,\sss} (kr,k\Delta\eta,\theta)=
\int^1_0 d\mu_k\,\frac{(1-\mu_k^2)^2}{2} J_0\left( A\sqrt{1-\mu_k^2}\right)\cos{\left(B\mu_k\right)}. \label{eq:HD_ss_ss}
\end{align}
As clear from the above two expressions, we expect that $\Gamma_{hh}$
and $\Gamma_{ss}$ show different $\theta$- and
$\Delta\eta$-dependences, which we confirm later. This property may
be useful in identifying the GW contribution from the noisy data that
contain the scalar contribution.

\begin{figure}[bt]
\begin{center}
\includegraphics[width=0.99\textwidth,angle=0,clip]{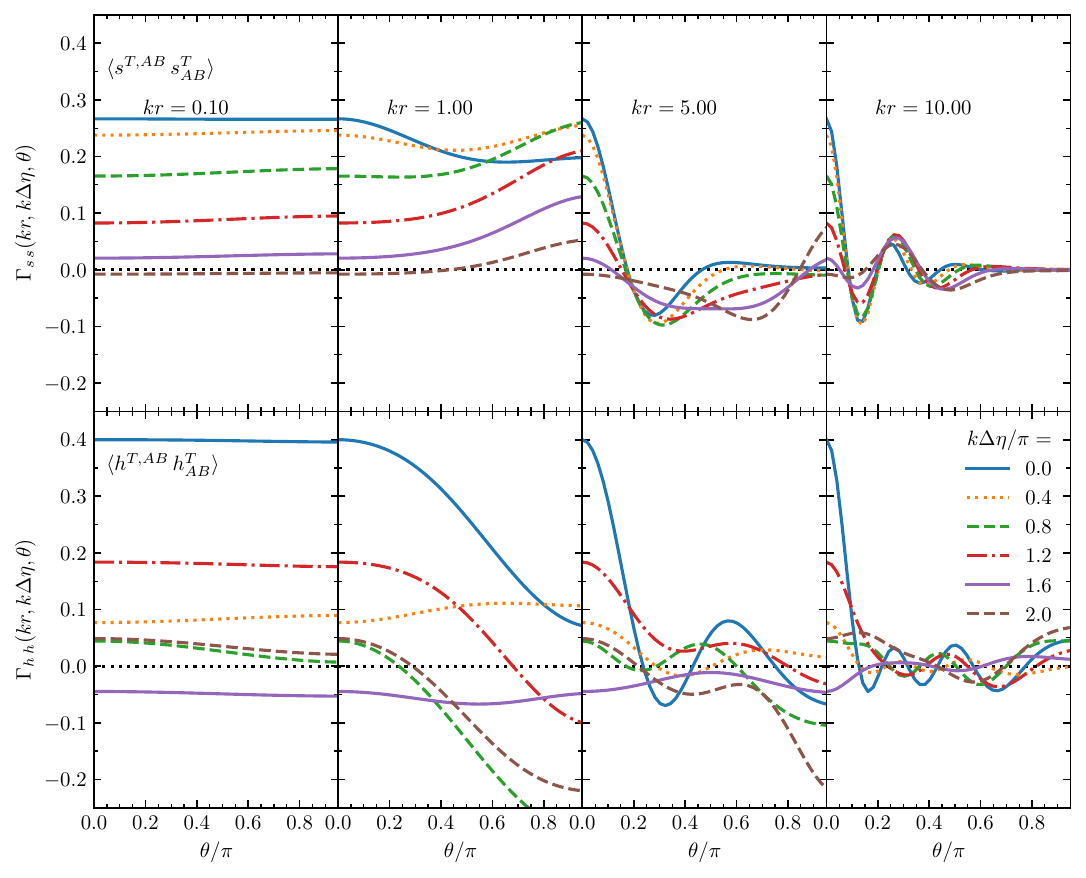}
\caption{ The kernels (ORFs) in the integrands of the correlation
  functions for the scalar and tensor (GW) components of the projected
  tidal force field.  The upper and lower rows show the ORFs for the
  scalar and tensor tidal fields, respectively.  Different columns
  show the results for several different values of $kr$ and different
  lines show those for different values of $k\Delta\eta=k(r'-r)$.  }
\label{fig:ORF_ss}
\end{center}
\end{figure}

Finally, let us present the correlation functions that contain only the GW contribution,
\begin{align}
\left\langle \ah^T  \etavx \ah^T \etapvxp
\right\rangle &= \frac{a_0^2}{a(\eta)a(\eta')}\int \frac{k^2 dk}{2\pi^2}P_h(k)\Gamma_{\ah\ah}(kx,k\Delta\eta,\theta), \\
\left\langle \ah^{T,A} \etavx \hh^T_A \etapvxp
\right\rangle &= \frac{a_0^2}{a(\eta)a(\eta')}\int \frac{k^2 dk}{2\pi^2}P_h(k)\chi(k)\Gamma_{\ah\,\hh}(kx,k\Delta\eta,\theta),
\end{align}
where
\begin{align}
\Gamma_{\ah\ah}(kx,k\Delta\eta,\theta)&=
\cos{(k\Delta\eta)}\int^1_0 d\mu_k\,\frac{\mu_k^2(1-\mu_k^2)^2}{4}J_0\left( A\sqrt{1-\mu_k^2}\right)\cos{\left(B\mu_k\right)},\label{eq:HD_ah_ah}
\\
\Gamma_{\ah\,\hh}(kx,k\Delta\eta,\theta)&=
\cos{(k\Delta\eta)}\int^1_0 d\mu_k\,\frac{\mu_k(1-\mu_k^4)}{4}J_0\left( A\sqrt{1-\mu_k^2}\right)\sin{\left(B\mu_k\right)}. \label{eq:HD_ah_hh}
\end{align}
Once again, we note that $\Gamma_{\ah\,\hh}$, is the important
quantity to extract parity-violating signals from the projected tidal
field in real space.  Since the cross-power spectrum is purely
imaginary as seen in eq.~(\ref{eq:power_proj_cros}), the ORF has
$\sin{(B\mu_k)}$ in the integrand rather than $\cos{(B\mu_k)}$.

\begin{figure}[bt]
\begin{center}
\includegraphics[width=0.99\textwidth,angle=0,clip]{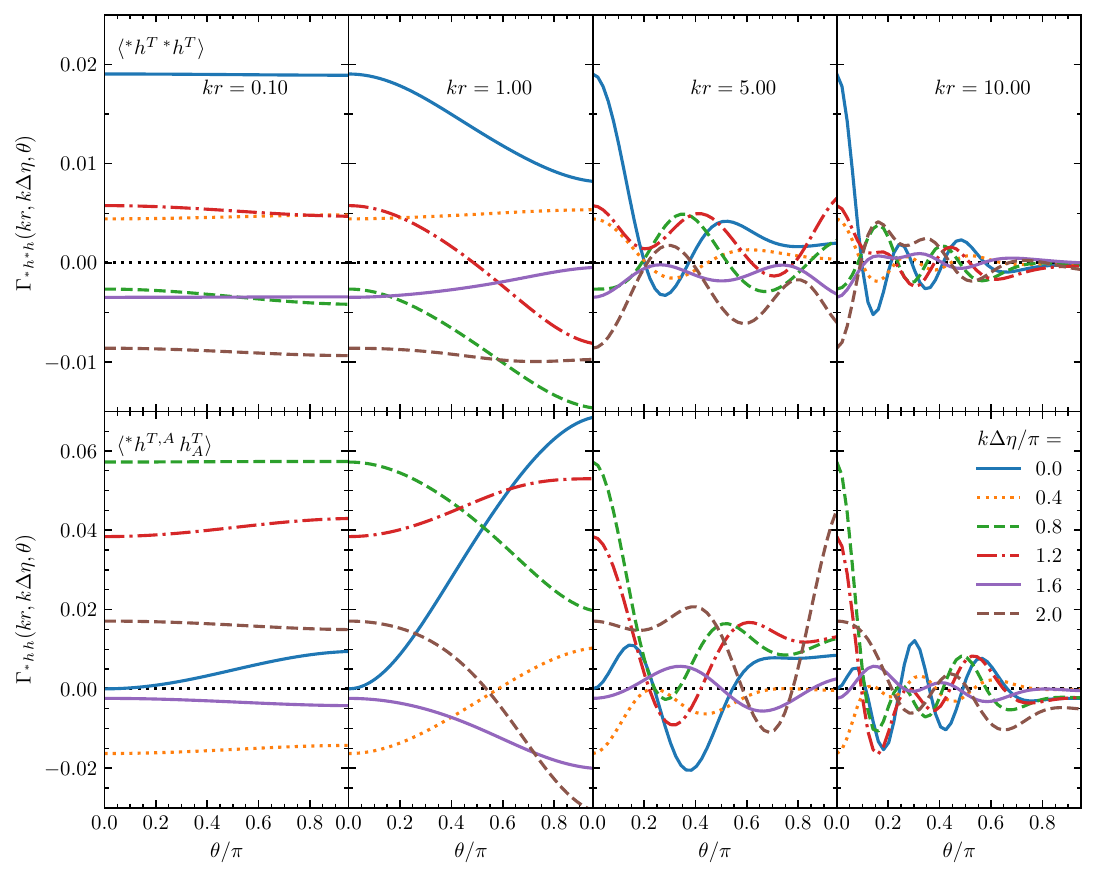}
\caption{ Similar with figure \ref{fig:ORF_ss} but for the
  auto-correlation function of the curl of the projected GWs ({\it
    upper row}) and its cross-correlation function with the divergence
  ({\it lower row}).  }
\label{fig:ORF}
\end{center}
\end{figure}

Figure \ref{fig:ORF_ss} shows the ORFs for the correlation functions
of the projected scalar ($\Gamma_{\sss\,\sss}$) and tensor
($\Gamma_{\hh\,\hh}$) tidal fields.  Both functions become almost
constant for varying $\theta$ at $kr\ll 1$ and fixed $k\Delta\eta$.
At $kr\gg 1$, both functions start to oscillate due to the term,
$\cos{(B\mu_k)} = \cos([-k \Delta \eta +
  2kr\sin^2{(\theta/2)}]\mu_k)$.  However, the ORF for the tensor
tidal field has the characteristic oscillatory feature due to the
factor $\cos(k\Delta\eta)$ because GWs propagate through spacetime,
while that of the scalar tidal field does not.  Thus, given an
accurate galaxy redshift survey data set, we confirm that it is in
principle possible to distinguish the contributions of the scalar and
tensor tidal forces by template matching applied to the projected
tidal field data in three dimensions.

Figure \ref{fig:ORF} shows the two ORFs for signals of GWs that can be
directly extracted from observation.  The upper row shows the result
for the divergence of the curl of the projected tidal tensor field,
$\Gamma_{\ah\ah}$. Although the overall behavior of the kernel
$\Gamma_{\ah\ah}$ is similar to that of $\Gamma_{\hh\,\hh}$, GW
signals can be directly extracted from this quantity since $\as \equiv
0$.  The lower row shows the ORF for parity violation signals of the
tidal tensor field. It becomes zero at the $\theta=0$ and
$k\Delta\eta=0$ limit, unlike the ORFs for the other
auto-correlations (see, e.g., Ref.\cite{Seto:2007} for a similar trend).

Finally let us comment on the corresponding quantities defined in
terms of the dimensionless tidal force field
$t_{ij}\approx(k^2/3\calH^2)h_{ij}$ instead of $h_{ij}$. Under the
assumption that the universe is matter-dominated, we have
$k^2/\calH^2\propto a(\eta)$. Hence this scale factor dependence
cancels the $1/a(\eta)$ dependence of $h_{ij}$.  This means that the
amplitude of $t_{ij}$ at the epoch of matter-radiation equality is
equal to that of $t_{ij}$ at any later times.  In the actual universe,
because of the dark energy contribution, this estimate is no longer
accurate, but it nevertheless gives a fairly good estimate of the
amplitude of $t_{ij}$.

\section{Conclusions and discussion}
\label{sec:conclusions}

Gravitational waves (GWs) are known to be imprinted into the tidal
force field as tensor perturbations. It can be estimated by defining
E-/B-modes and computing their power spectra. In this paper, we have
developed a new technique to estimate the contribution of GWs in the
projected tidal field.  We introduced two two-dimensional vector
fields in three dimensions constructed by taking the divergence and
curl of the projected tidal field, denoted by $h^A$ and $\ah^A$,
respectively. The auto-power spectra naturally contain contributions
of the scalar-type tidal field.  We found that further taking the
divergence of $\ah^A$, which is a pseudo-scalar quantity, enable us to
remove the scalar contribution and thus to single out the GW
contribution.  The cross-correlation between $h^A$ and $\ah^A$ is free
from the scalar tidal contribution, and it is non-vanishing only if
parity is violated. Thus signals of parity violation in the GW
background can be extracted from this cross-correlation, if any.  It
roughly corresponds to measuring the cross-power spectrum of $E$ and
$B$-modes, but these are complementary to each other in the sense that
our estimator can be naturally defined locally in position space.  We
expressed the two-point functions in the Fourier integral form where
the integrand is divided to the two factors; the power spectrum and
the kernel, which we called the overlap reduction function (ORF),
following the terminology used in pulsar timing array data
analysis. We pointed out that the difference between the ORF of the
tensor component and that of the scalar component may be used to
identify the GW contribution in the tidal force field.

We now discuss how the tidal force tensor studied above is imprinted
into the observed galaxy shapes.  The scalar tidal field is known to
be linearly related to the galaxy shape, known as the linear alignment
model \cite{Catelan:2001,Hirata:2004,Blazek:2011,Okumura:2019}.  Under
this model the observed projected halo shape field, $\GG^{S}_{AB}$, is
related to the projected tidal field $\sss_{AB}$ linearly as $
\GG^{S}_{AB}\etavk = b_K^{\rm S}(\eta) \sss_{AB}\etavk $.  We can
observe only the trace-free part of the projected galaxy shape. Here
thus we omit the superscript $T$ from all the quantities for clarity.
The coefficient $b_K^{\rm S}$ is called the linear shape bias for the
scalar tidal field, in analogy with the linear galaxy bias,
$\delta_g\etavk=b(\eta)\delta_m\etavk$.  Ref.~\cite{Schmidt:2014}
proposes an ansatz that the effect of gravitational waves is also
imprinted into the shape of galaxies at linear order,
\begin{align}
\GG^{\rm GW}_{AB}\etavk=b_t^{\rm GW}\etak t_{AB}\etavk, \label{eq:gamma_la_t}
\end{align}
and the total shape field of galaxies is expressed as the sum of the
scalar and tensor contributions
\cite{Schmidt:2014,Biagetti:2020,Akitsu:2023}
\begin{align}
\gamma_{AB}\etavk = \gamma_{AB}^S\etavk+\gamma_{AB}^{\rm GW}\etavk = b_K^S(\eta)\sss_{AB}\etavk + b_t^{\rm GW}\etak t_{AB}\etavk. 
\end{align}
The linear shape bias for GWs, $b_t^{\rm GW}$, is not independent of
$b_K^{\rm S}$. By matching the second-order density induced by scalar
and tensor tidal fields, Ref.~\cite{Schmidt:2014} introduced the
ansatz for the relation between them as $b_t^{\rm GW}\etak \propto
\alpha\etak b_K^{\rm S}(\eta)$ (see
Refs.~\cite{Schmidt:2014,Akitsu:2023} for the explicit form of
$\alpha$).  This relation has been tested and confirmed using $N$-body
simulations \cite{Akitsu:2023}.  Using this relation, we can
immediately obtain the power spectrum of projected shape from that of
the projected tidal field derived in section
\ref{sec:power_projected}.  By taking the divergence and curl
operation on the galaxy shape, we can obtain $\GG_A\etavx,
\ag_A\etavx$ and $\ag\etavx$ as $\GG_A = \partial^B \GG_{BA}, \ag_A =
\epsilon^{BC}\partial_B\GG_{CA}$, and $\ag = \partial^A\,\ag_A$,
respectively. Note again that all these quantities are trace-free and
the superscript $T$ is omitted. Since they are linearly related to the
projected tidal force field, measuring the auto-correlation of $\ag$
enables us to extract GW signals. The cross-correlation of the
divergence ($\GG_A$) and curl ($\ag_A$) of the projected shape field
becomes nonzero only if there exists parity-violating signals.

In this paper we considered various types of correlations of the
projected tidal force field in three dimensions, which are observable
in spectroscopic galaxy surveys. The positions of galaxies are sampled
by their redshifts in such galaxy surveys, and thus affected by
redshift-space distortions (RSD) \cite{Kaiser:1987}.  We assumed the
validity of linear theory.  While the shape field is not affected by
RSD on linear scales, it is on nonlinear scales
\cite{Matsubara:2024,Okumura:2024}.  Furthermore, on such small
scales, the linear relation between the tidal field and galaxy shapes
needs to be modulated. We thus need to consider the higher-order
nonlinear shape bias effect \cite[e.g.,][]{Akitsu:2023a}.  To place
quantitative constraints on the amplitude of GWs and the chiral
parameter, we need to consider more accurate models. It is left to our
future work.

\begin{acknowledgments}
We thank Kazuyuki Akitsu, Atsushi Taruya and Toshiki Kurita for fruitful discussion.  We also thank the
workshop
\href{https://events.asiaa.sinica.edu.tw/workshop/20231204/}{`\emph{Large-scale
  Parity Violation Workshop}'} held at Academia Sinica Institute of
Astronomy and Astrophysics (ASIAA) during which this project was
advanced.  T.O. acknowledges support from the Ministry of Science and
Technology of Taiwan under grants No. NSTC 112-2112-M-001-034- and No. NSTC 113-2112-M-001-011- and the Career Development Award,
Academia Sinica (AS-CDA-108-M02) for the period of 2019-2023.  This
work is also supported in part by JSPS KAKENHI Nos.~JP20H05853 and
JP24K00624.
\end{acknowledgments}

\appendix

\section{Relation to E/B-mode decomposition}\label{sec:EB}

We describe how our formalism is related to that of the E/B-modes
shown in Ref.~\cite{Akitsu:2023,Philcox:2024,Saga:2024}. The
2-dimensional metric perpendicular to the line of sight, $g_{AB}$, is
given by
\begin{align}
g_{AB} = \hat{x}_A \hat{x}_B + \hat{y}_A \hat{y}_B  \, .
\end{align}
We also define the polarization tensors of the E- and B-modes,
respectively as
\begin{align}
e^{(E)}_{AB} = \frac{1}{\sqrt{2}} \left(\hat{x}_A \hat{x}_B - \hat{y}_A \hat{y}_B\right), \quad
e^{(B)}_{AB} = \frac{1}{\sqrt{2}} \left(\hat{x}_A \hat{y}_B + \hat{y}_A \hat{x}_B\right).
\end{align}
These E- and B-modes are related to the R- and L-mode polarization
tensors used in the main text as
\begin{align}
\big[ e_{AB}^{(R,L)} \big]^T &=  e_{AB}^{(R,L)}-\frac12 \big( g^{CD}e_{CD}^{(R,L)} \big) \delta_{AB} 
=\frac{1}{\sqrt{2}}\left[ \frac{1+\mu_k^2}{2} e_{AB}^{(E)} \pm i\mu_k e_{AB}^{(B)} \right],
\end{align}
where $\mu_k$ is the directional cosine of the wavevector $\vk$ with
respect to the line of sight, and the superscript $T$ stands for the
quantity with the trace subtracted.  Then the traceless $[h_{AB}]^T$
reads:
\begin{align}
\big[ h_{AB} \big]^T &= h_{(R)} \big[ e_{AB}^{(R)} \big]^T + h_{(L)} \big[ e_{AB}^{(L)} \big]^T 
\nn \\ &
= \frac{1}{\sqrt{2}}\left[ \frac{1+\mu_k^2}{2} e_{AB}^{(E)} + i\mu_k e_{AB}^{(B)} \right] h_{(R)}
-
\frac{1}{\sqrt{2}}\left[ \frac{1+\mu_k^2}{2} e_{AB}^{(E)} - i\mu_k e_{AB}^{(B)} \right] h_{(L)}\end{align}
Thus, the E- and B-modes of the projected GWs are expressed as
\begin{align}
h_{(E)} &= h_{AB} e^{AB}_{(E)} = \frac{1}{\sqrt{2}}\frac{1+\mu_k^2}{2}(h_{(R)}+h_{(L)}), \\
h_{(B)} &= h_{AB} e^{AB}_{(B)} = \frac{i}{\sqrt{2}}\mu_k(h_{(R)}-h_{(L)}).
\end{align}
Finally, the power spectra $P_{XY}$ defined by $\big\langle
h^{(X)}\etavkp h_{(Y)}\etapvkp\big\rangle=\deltakk P_{XY}\etaetapvk$,
where $XY=\{EE,BB,EB\}$, are respectively given by
\begin{align}
P_{EE}\etaetapvk &= \frac18 (1+\mu_k^2)^2 (P_{(R)}+P_{(L)})= \frac18 (1+\mu_k^2)^2 P_{h}\etaetapk, \\
P_{BB}\etaetapvk &= \frac14 \mu_k^2 (P_{(R)}+P_{(L)})= \frac14 \mu_k^2 P_{h}\etaetapk, \\
P_{EB}\etaetapvk &= \frac{i}{4} \mu_k (1+\mu_k^2)(P_{(R)}-P_{(L)})= \frac{i}{4} \mu_k (1+\mu_k^2)\chi(k)P_{h}\etaetapk,
\end{align}
These are the power spectra already given in
Refs.~\cite{Akitsu:2023,Philcox:2024,Saga:2024}.

\section{Power spectra of projected field before subtracting the trace}\label{sec:power_w_trace}

The tidal force field corresponding to observables is the projected
field with the trace part subtracted, and the expressions of the power
spectra were shown in section \ref{sec:projection}. In this appendix,
for completeness, we present the power spectra of the projected field
before the trace part is subtracted, though they may not be direct
observables.

The power spectrum of the projected scalar tidal field is given by  
\begin{align}
\left\langle  \sss^{AB}\etavk\sss_{AB}\etapvkp\right\rangle 
=k^4 \deltakk(1-\mu_k^2)^2 P_\psi \etaetapk \, . \label{eq:sabsab}
\end{align}
By comparing this with equation (\ref{eq:sTabsTab}), one can see that
the shape of the power spectrum for the projected scalar tidal field
is unchanged,
$\left\langle \sss^{AB}\sss_{AB}\right\rangle 
=2\left\langle \sss^{T,AB}\sss^T_{AB}\right\rangle$.
The power spectrum of the projected tensor tidal field without
subtracting the trace part is given by
\begin{align}
\left\langle \hh^{AB}\etavk\hh_{AB}\etapvkp\right\rangle
&=\frac14\deltakk(1+\mu_k^2)^2 P_h\etaetapk. \label{eq:habhab} 
\end{align}

The divergence and curl of the projected field before the trace is
subtracted are given by equation (\ref{eq:XA_aXA}). The power spectrum
of the divergence is given by
\begin{align}
\left\langle \sss^{A}\etavk\sss_A\etapvkp \right\rangle
=
\deltakk
k^6 (1-\mu_k^2)^3 P_\psi\etaetapk, \label{eq:sasa}
\end{align}
where it is related to that with the trace part subtracted via
$\left\langle \sss^{A}\sss_A \right\rangle = 4\left\langle
\sss^{T,A}\sss^T_A\right\rangle $. Since $\as_A=0$, $\left\langle
\as^{A}\, \as_A \right\rangle =0$.

The auto-power spectra of the divergence and curl of the projected
tidal tensor field are, respectively, given by
\begin{align}
& \left\langle \hh^A\etavk\hh_A\etapvkp \right\rangle  
= 
\frac{k^2}{4}\deltakk
\mu_k^2(1-\mu_k^4) P_h\etaetapk, \label{eq:power_proj_auto_h}
\\
& \left\langle \ah^A\etavk\ah_A\etapvkp \right\rangle  
= 
\frac{k^2}{4}\deltakk
(1-\mu_k^4) P_h\etaetapk. \label{eq:power_proj_auto}
\end{align}
These two power spectra are related to each other as
$\big\langle \hh^A\hh_A\big\rangle =\mu_k^2  \big\langle \ah^A \ah_A\big\rangle$.

Finally, the two power spectra, which can directly probe a GW signal
and parity-violating signals, have the same forms before and after the
trace part subtracted:
\begin{align}
& \left\langle \ah^{T}\etavk\ah^T\etapvkp \right\rangle  
= \left\langle \ah\etavk\ah\etapvkp \right\rangle \, , \\  
& \left\langle \ah^{T,A}\etavk\hh^T_A\etapvkp \right\rangle  
= 
\left\langle \ah^{A}\etavk\hh_A\etapvkp \right\rangle  \, .
\end{align}

\section{Derivation of power spectra of projected tidal force field} \label{sec:derivation}
\subsection{Power spectra of scalar tidal field}\label{sec:derivation_scalar}
In this appendix, we provide derivations for the formulas of the power spectra of the scalar tidal field presented in section \ref{sec:projection}. 
The Fourier component of the projected scalar tidal field with the trace component subtracted, $s_{AB}^T\etavx$, is given by \begin{align}
s_{AB}^T\etavk=\left(-k_Ak_B+\frac12\delta_{AB}k^Ck_C\right)\psi\etavk. \label{eq:sTAB}
\end{align}
Thus it is straightforward to obtain the auto-power spectrum (equation (\ref{eq:sTabsTab})), 
\begin{align}
&\left\langle s^{T,AB}\etavk s_{AB}^T\etapvkp\right\rangle
\nn \\ & \quad 
=
\left(-k^Ak^B+\frac12\delta^{AB}k_Ck^C\right)
\left(-k'_Ak'_B+\frac12\delta_{AB}k'_Dk'^D\right) 
\left\langle \psi\etavk\psi\etapvkp\right\rangle 
\nn \\ &  \quad 
= k^4\left[ (1-\mu_k^2)^2 - (1-\mu_k^2)^2+\frac12(1-\mu_k^2)^2 \right]\left\langle \psi\etavk\psi\etapvkp\right\rangle 
\nn \\ & \quad
=\frac12 \deltakk k^4 (1-\mu_k^2)^2 P_\psi \etaetapk \, .
\end{align}

For the divergence and curl of the projected scalar tidal field with the trace part subtracted, we have from equation (\ref{eq:sTAB}), respectively,
\begin{align}
\sss^T_A\etavk=-\frac{i}{2}k^Bk_Bk_A\psi\etavk, \qquad
\as^T_A\etavk=\frac{i}{2}\epsilon^{CA}k^Bk_Bk_C\psi\etavk.
\end{align}
The auto-power spectrum of the latter, shown in equation (\ref{eq:sTsT}), is obtained as 
\begin{align}
&
\left\langle \as^{T,A}\etavk \as^T_A\etapvkp \right\rangle 
\nn \\ & \quad 
=-\frac14 \left(\epsilon^{CA}k^Bk_Bk_C\right)\left(\epsilon_{EA}k'^Dk'_Dk'^E\right) \left\langle  \psi\etavk\psi\etapvkp \right\rangle 
\nn \\ & \quad
=-\frac14 (\delta^{C}_{E}\delta^{A}_{A}-\delta^{C}_{A}\delta^{E}_{A})(k^Bk_B)(k'^Dk'_D)k_Ck'_E\left\langle  \psi\etavk\psi\etapvkp \right\rangle 
\nn \\ & \quad
= \frac14 \deltakk
k^6 (1-\mu_k^2)^3 P_\psi\etaetapk \ .
\end{align}
The auto-power spectrum of $\sss^T_A$ can be derived similarly and it is easy to show that they are equivalent, 
$\left\langle \sss^{T,A}\etavk \sss^T_A\etapvkp \right\rangle
=\left\langle \as^{T,A}\etavk \as^T_A\etapvkp \right\rangle
$.

Note that when we perform the calculations for the scalar tidal field without subtracting the trace part, we can instead obtain equations (\ref{eq:sabsab}) and (\ref{eq:sasa}).

\subsection{Power spectra of tensor tidal field} \label{sec:derivation_tensor}
Here, we provide derivations for the formulas of the power spectra of the tensor tidal field presented in section \ref{sec:projection}. 
Throughout this appendix, we repeatedly use the relation,
\begin{align}
h\indices{^z_z} = -\sum_{\lambda=R,L}
e_{AA}^{(\lambda)}
h_{(\lambda)} = -\frac12(1-\mu_k^2)\sum_{\lambda=R,L}h_{(\lambda)}. 
\end{align}
Since GWs are traceless in three dimensions, 
$
h\indices{^A_A}=h\indices{^x_x}+h\indices{^y_y} = -h\indices{^z_z} $.

The traceless part of the projected GWs is given by 
\begin{align}
h^T_{AB}\etavk=h_{AB}\etavk+\frac12\delta_{AB}h\indices{^z_z}\etavk. \label{eq:hTAB}
\end{align}
The temporal auto correlation function in Fourier space is obtained as
\begin{align}
&\left\langle h^{T,AB}\etavk h_{AB}^T\etapvkp\right\rangle 
\nn \\ & \qquad
=
\left\langle \left[h^{AB}\etavk+\frac12\delta^{AB}h\indices{^z_z}\etavk \right]
\left[h_{AB}\etapvkp+\frac12\delta_{AB}h\indices{^z_z}\etapvkp \right]\right\rangle 
\nn \\ & \qquad
= \left\langle  h^{AB}\etavk h_{AB}\etapvkp \right\rangle
+ \left\langle  h\indices{^A_A}\etavk h\indices{^z_z}\etapvkp \right\rangle
+\frac12 \left\langle  h\indices{^z_z}\etavk h\indices{^z_z}\etapvkp \right\rangle 
\nn \\ & \qquad
=\frac18\deltakk(1+6\mu_k^2+\mu_k^4) P_h\etaetapk \, ,
\end{align}
which gives equation (\ref{eq:hTabhTab}). 
Here we used the relation,
$\sum_{\lambda=R,L}{e^{AB}_{(\lambda)} \bar{e}_{AB}^{(\lambda)}} = (1+\mu_k^2)^2/2$.

In the rest of this appendix, we present derivations of the power spectra of divergence and curl of the projected tidal tensor field given in section \ref{sec:power_div_curl}. Let us first show the explicit form of the divergence and curl of the projected tidal tensor field, $\hh_A$ and $\ah_A$, and the divergence of the curl of the field, $\ah$. We start by showing the expressions before subtracting the trace part. They are given by
\begin{align}
\hh_A\etavk &=ik^Bh_{BA}\etavk= ik^B \sum_{\lambda=R,L} e^{(\lambda)}_{BA}(\hat\vk)h_{(\lambda)}\etavk 
\nn \\& 
= -\frac{k}{2}\cos\theta_k \sin\theta_k \left[ (h_{(R)}-h_{(L)})\hat x_A + i\cos{\theta_k} (h_{(R)}+h_{(L)})\hat y_A\right], \label{eq:h_A}
\\ 
\ah_A\etavk &=ik_B\epsilon^{BC}h_{CA}\etavk= ik_B \epsilon^{BC}\sum_{\lambda=R,L} e^{(\lambda)}_{CA}(\hat\vk)h_{(\lambda)}\etavk
\nn \\ &
= \frac{k}{2}\sin\theta_k \left[\cos\theta_k (h_{(R)}-h_{(L)})\hat y_A - i (h_{(R)}+h_{(L)})\hat x_A\right] \, , \label{eq:ah_A}
\\
\ah\etavk & = ik^A\ah_A\etavk
=\frac{i}{2}k^2(1-\mu_k^2)\mu_k(h_{(R)}-h_{(L)}) \, .
\end{align}
For the second equality in equation (\ref{eq:h_A}), we used the relations:
\begin{align}
\hat k^B e_{BA}^{(+)} &= \frac{1}{\sqrt{2}}\hat k^B 
\left( e^{(1)}_B e^{(1)}_A - e^{(2)}_B e^{(2)}_A \right) = -\frac{1}{\sqrt{2}}\cos^2\theta_k \sin\theta_k \hat y_A \, , \nn \\ 
\hat k^B e_{BA}^{(\times)} &= \frac{1}{\sqrt{2}}\hat k^B 
\left( e^{(1)}_B e^{(2)}_A + e^{(2)}_B e^{(1)}_A \right) = \frac{1}{\sqrt{2}}\cos\theta_k \sin\theta_k \hat x_A \, , \nn 
\end{align}
and similarly for the second equality in equation (\ref{eq:ah_A}), we used the relations:
\begin{align}
\hat k_B \epsilon^{BC} e_{CA}^{(+)} &= -\frac{1}{\sqrt{2}}\sin\theta_k \hat x_A \, , 
\qquad 
\hat k_B \epsilon^{BC} e_{CA}^{(\times)} = -\frac{1}{\sqrt{2}}\cos\theta_k \sin\theta_k \hat y_A \, . \nn 
\end{align}

Next we show the divergence and curl of the projected tidal tensor field with the trace part subtracted, namely, $\hh^T_A$, $\ah^T_A$ and $\ah^T$. 
From equation (\ref{eq:hTAB}), we have
\begin{align}
\hh^T_A\etavk &= \hh_A\etavk + \frac{i}{2}k_A h\indices{^z_z}\etavk, 
\nn \\ &
= -\frac{k}{2}\sin\theta_k \left[\cos\theta_k (h_{(R)}-h_{(L)})\hat x_A + \frac{i}{2}(1+\cos^2{\theta_k}) (h_{(R)}+h_{(L)})\hat y_A\right],
\\
\ah^T_A\etavk &= \ah_A\etavk
+\frac{i}{2}k^B \epsilon_{BA}h\indices{^z_z}\etavk 
\nn \\ &
= \frac{k}{2}\sin\theta_k \left[\cos\theta_k (h_{(R)}-h_{(L)})\hat y_A - \frac{i}{2}(1+\cos^2{\theta_k}) (h_{(R)}+h_{(L)})\hat x_A\right]
\\
\ah^T\etavk &= ik^A\ah^T_A\etavk 
= \ah\etavk
\, . 
\end{align}
From the above, one can derive the temporal correlation functions of the various quantities:
\begin{align}
    \left\langle \hh^{T,A}\etavk \hh^T_A\etapvkp \right\rangle 
    & 
    = \left\langle \left( h^{A}\etavk + \frac{i}{2}k^A h\indices{^z_z}\etavk \right)  \left( h_{A}\etapvkp + \frac{i}{2}k_A h\indices{^z_z}\etapvkp \right) \right\rangle
    \nn \\ & 
    =\frac{1}{4}\deltakk\sin^2\theta_k \left[ \cos^2\theta_k +\frac14(1+\cos^2\theta_k)^2\right]  P_h\etaetapk
    \nn \\ & 
= \frac{1}{16}\deltakk k^2(1-\mu_k^2)(1+6\mu_k^2+\mu_k^4) P_h\etaetapk \, , 
 \\
 ~\nn\\
 \left\langle\ah^T\etavk\ah^T\etapvkp\right\rangle 
&=\left\langle \left(ik^A \ah_{A}\etavk\right)\left(ik^B \ah_{B}\etapvkp\right) \right\rangle\nn \\
&
=\frac{1}{4}\deltakk
k^4\mu_k^2(1-\mu_k^2)^2 P_h\etaetapk 
\, ,
\\
 ~\nn\\
 \left\langle\ah^{T,A}\etavk\hh^T_A\etapvkp\right\rangle 
&=\left\langle \left( \ah^A\etavk + \frac{i}{2}k_B\epsilon^{BA}h\indices{^z_z}\etavk\right)\left(h_{A}\etapvkp + \frac{i}{2}k_A h\indices{^z_z}\etapvkp\right) \right\rangle\nn \\
&
=\frac{i}{4}\deltakk 
k^2\cos\theta_k \sin^2\theta_k (\cos^2\theta_k + 1) P_h\etaetapk 
\nn \\
&
=\frac{i}{4}\deltakk
k^2\mu_k(1-\mu_k^4) P_h\etaetapk 
\, ,
\end{align}
which respectively give equations (\ref{eq:power_proj_auto_h_traceless}), (\ref{eq:power_proj_auto_h_T}) and (\ref{eq:power_proj_cros}).

\bibliography{ms_jcap.bbl}
 \end{document}